\documentclass[aps,prl,reprint,superscriptaddress,showpacs,longbibliography,
floatfix,footnoteinbib]{revtex4-1}
\usepackage{amsmath,amssymb,graphicx,units,ulem}
\usepackage[plainpages=false,pdfpagelabels,colorlinks=true,linkcolor=red,
urlcolor=blue,citecolor=blue,pdftitle={Pairing and superconductivity
  in the flat band: Creutz lattice},pdfauthor={},
pdfdisplaydoctitle=true,pdfduplex=DuplexFlipLongEdge]{hyperref}

\begin{document}

\title{Pairing and superconductivity in the flat band: Creutz lattice}

\author{Rubem Mondaini} 
\affiliation{Beijing Computational Science Research Center, Beijing
  100193, China}
\author{G. G. Batrouni}
\affiliation{Universit\'e C\^ote d'Azur, INPHYNI, CNRS, 0600 Nice,
  France}
\affiliation{MajuLab, CNRS-UCA-SU-NUS-NTU International Joint Research
  Unit, 117542 Singapore}
\affiliation{Centre for Quantum Technologies, National University of
  Singapore, 2 Science Drive 3, 117542 Singapore}
\affiliation{Beijing Computational Science Research Center, Beijing
  100193, China}
\author{B. Gr\'emaud}
\affiliation{MajuLab, CNRS-UCA-SU-NUS-NTU International Joint Research
  Unit, 117542 Singapore}
\affiliation{Centre for Quantum Technologies, National University of
  Singapore, 2 Science Drive 3, 117542 Singapore}
\affiliation{Department of Physics, National University of Singapore, 2
  Science Drive 3, 117542 Singapore}
\affiliation{Laboratoire Kastler Brossel, UPMC-Sorbonne Universit\'es,
  CNRS, ENS-PSL Research University, Coll\'ege de France, 4 Place
  Jussieu, 75005 Paris, France} 
\affiliation{Aix Marseille Universit\'e, Universit\'e de Toulon, CNRS, CPT, Marseille, France}

\begin{abstract}
We use unbiased numerical methods to study the onset of pair
superfluidity in a system that displays flat bands in the
noninteracting regime. This is achieved by using a known example of
flat band systems, namely the Creutz lattice, where we investigate the
role of local attractive interactions in the $U<0$ Hubbard
model. Going beyond the standard approach used in these systems where
weak interactions are considered, we map the superfluid behavior for a
wide range of interaction strengths and exhibit a crossover between
BCS and tightly bound bosonic fermion pairs. We further contrast these
results with a standard two-leg fermionic ladder, showing that the
pair correlations, although displaying algebraic decay in both cases,
are longer ranged in the Creutz lattice, signifying the robustness of
pairing in this system.
\end{abstract}


\maketitle

\section{Introduction}
\label{intro}

Systems exhibiting flat (dispersionless) bands come in many varieties
and manifest a wide range of interesting phenomena such as exotic
superfluid phases, edge states, topological insulator/superconductor
phases, and bound Majorana fermion edge states to name a few. For
example, at half filling in the Lieb lattice (which belongs to a large
family of flat band models)
\cite{mielke92,aoki96,deng03,wu07,tasaki08,lan12,jacqmin}, the
fermionic Hubbard model with repulsive contact interaction, $U$, has a
ground state with nonzero spin~\cite{lieb89}, while in the absence of
Hubbard interaction, a particle in the flat band is geometrically
localized on four sites due to quantum interference in the hopping
terms~\cite{sutherland86}. On the other hand, the attractive Hubbard
model on the same lattice exhibits unusual charge and charge transfer
signatures within the flat band and reduced pairing order when either
the flat band starts to be occupied or when it is completely
filled~\cite{iglovikovv14}. It has also been argued that fermionic
pairing in flat bands would lead to more robust pairs and higher
critical temperatures~\cite{heikkila16}. Remarkable experiments have
recently shown~\cite{Cao2018} that graphene bi-layers twisted by about 
1.1$^\circ$ exhibit an ultra-flat band at the charge neutrality
point. This leads to a correlated insulator which, when doped, becomes
superfluid.

Another very interesting class of systems has the lowest band flat,
 e.g., sawtooth~\cite{huber10}, kagome, Creutz~\cite{Creutz1999}, and
many others~\cite{misumi17}. In the ground state of such systems,
there is a critical density below which each particle is geometrically
localized over a few sites (which depends on the lattice geometry) and
such that the localized particles do not interact. Exceeding the
critical density causes the particle wavefunctions to overlap and
interaction ensues, thus destroying localization. Such systems have
been recently studied extensively both in fermionic and bosonic
models. In the latter case, it was shown in a fully
  frustrated chain (diamond lattice, which has three flat bands when
  threaded by a $\pi$ flux) that when interactions are included in
  such a way to reduce the original local U(1) symmetry to a
  discrete local $Z_2$ gauge symmetry, a new exotic phase appears,
  namely a nematic superfluid where the current is supported by bosons
  paired on different sites\cite{doucot02}, or using a fermionic
  language by the condensation of pairs of Cooper pairs~\cite{Rizzi06};
  the same system was later studied for spinless
  fermions~\cite{lopes11}. In the sawtooth lattice, it was found that
doping above the critical density leads to a phase where the peak of
the momentum distribution is at nonzero momentum~\cite{huber10}, while
doping the kagome lattice leads to a supersolid
phase~\cite{huber10}. Adding longer range interactions between the
bosons on the sawtooth lattice uncovers topological effects such as
the Haldane insulator phase and edge states in open
systems~\cite{gremaud17}. Bosons were also studied on the flat band
Creutz lattice (Fig.~\ref{fig:creutzlattice}) leading to a rich phase
diagram exhibiting a condensate, a pair condensate, a supersolid and
phase separation phases~\cite{Takayoshi2013,Tovmasyan2013}.

Fermions in such systems, where the flat band is the lowest, are very
interesting for a variety of reasons. They have been shown to exhibit
a plethora of topological effects such as edge states~\cite{Creutz1999},
which are robust to interactions when in the presence of induced
pairing terms~\cite{sticlet14}.  A general treatment of fermions with
attractive interactions in nontrivial flat bands, where the Chern
number ${\cal C}\neq 0$, demonstrated~\cite{peotta15} that such
systems are guaranteed to exhibit nonzero superfluid weight $D_s\geq
|{\cal C}|$. In the case of quasi-one dimensional systems, in
particular the Creutz lattice, with ${\cal C}=0$, it was shown
~\cite{Tovmasyan2016} that the superfluid weight is $D_s \geq |{\cal
  W}|^2$, where ${\cal W}$ is the winding number.
  
Here, we study numerically the Creutz model with an attractive Hubbard
interaction, $U<0$, using the density matrix renormalization group
(DMRG) and exact diagonalization (ED). We calculate the superfluid
weight, the Drude weight $D_s$ in one dimension, for various fillings
as a function of $|U|$ and show that for small $|U|$ the predictions
of Ref.~\onlinecite{Tovmasyan2016} are accurate, \textit{i.e.} the
superfluid weight grows linearly with the strength of the interactions
for densities away from half-filling. Going beyond the small $|U|$
regime, the model is shown to map onto a hardcore bosonic model for
large enough $|U|$. As a measure of the robustness of superfluidity,
we calculate the one- and two-particle gaps, and the decay exponents
of the pair correlation functions for the Creutz flat band model and
compare with the normal two-leg model. We find that pair correlations
decay algebraically with distance, whereas single-particle ones decay
exponentially, for finite values of the interactions. 
  Furthermore, we find that for all the parameters we studied, the
  power-law decay is slower, often much slower, in the Creutz lattice
  than in the normal two-leg model.

\section{Model}
\label{modelsection}

\begin{figure}[h!]
\includegraphics[width=0.8\columnwidth]{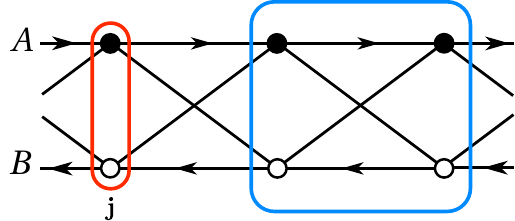}
\caption{(Color online) The Creutz lattice~\cite{Creutz1999}. The
  (red) rectangle encloses a unit cell, the (blue) square shows the
  geometry of the localized states in the absence of interactions when
  the particle density is less than the critical value $1/2$. The
  arrows depict the sign of the hopping in the intrachain bonds. In
  the interchain bonds, the hopping energies have the same magnitude.}
\label{fig:creutzlattice}
\end{figure}

We study the attractive Hubbard model on the Creutz
lattice~\cite{Creutz1999} (see Fig.~\ref{fig:creutzlattice}), governed
by the Hamiltonian
\begin{eqnarray}
\nonumber
{\cal H} &=&-{\rm i}t\sum_{j,\sigma} \left (c^{A\dagger}_{j,\sigma}
c^{A}_{j+1,\sigma} -c^{B\dagger}_{j,\sigma}
c^{B}_{j+1,\sigma} + {\rm H.c.}\right )\\
\nonumber
&&-t\sum_{j,\sigma} \left ( c^{A\dagger}_{j,\sigma}
c^{B}_{j+1,\sigma} + c^{B\dagger}_{j,\sigma}
c^{A}_{j+1,\sigma} + {\rm H.c.}\right )\\
&& + U \sum_{j,\alpha} 
  n^\alpha_{j,\uparrow}n^\alpha_{j,\downarrow},
\label{eq:creutzham}
\end{eqnarray}
where the onsite interaction, $U$, is negative; $A$ and $B$ label the two
chains, $t$ connects both inter-
and intra-chain sites $j$ and $j+1$ and the sum over $j$ spans $L$
unit cells. The fermion spin is labeled by
$\sigma=\uparrow,\,\downarrow$ and $\alpha=A,B$ is the chain
index. This Hamiltonian governs a balanced population of up and down
spins making inter- and intra-chain hops and interacting attractively
when on the same site.

It is worth noting that applying a local gauge transformation,
$c_{j,\sigma}^{A} \to {\rm exp}(-{\rm i}\pi j/2)c_{j,\sigma}^{A} $ and
$c_{j,\sigma}^{B} \to {\rm exp}(-{\rm i}\pi (j-1)/2)c_{j,\sigma}^{B} $
, renders all the hopping terms real. The intra-chain hopping
parameter on chain $A$ ($B$) becomes $t$ ($-t$); inter-chain hopping
between site $j$ on chain $A$ ($B$) and site $j+1$ on chain $B$ ($A$)
is given by $-t$ ($t$). When applied to a lattice with periodic
boundary conditions, the number of unit cells must be a multiple of
four for this transformation to apply, whereas for the case of open
boundary conditions, it works for any system size. This proved to be
very useful in some of our DMRG calculations on large lattices, as one
has to deal with a purely real Hamiltonian. One must note
  that, in general, local gauge transformations do not change the
  topological class of the system as long as the relevant symmetries
  are consistently modified when changing gauge~\cite{Ryu2010}.

For $U=0$, ${\cal H}$ can be
diagonalized~\cite{Creutz1999,Takayoshi2013,Tovmasyan2013} revealing
two flat bands, $E_\pm = \pm 2t$. Consequently, when a particle is
placed in the lattice, it will be localized on four sites, shown by
the blue square in Fig.~\ref{fig:creutzlattice}. The localized ground-states are given by,
\begin{equation}
|\Psi^{loc}_{j,\sigma}\rangle =- \frac{1}{2}\left
[c^{B\dagger}_{j,\sigma}
  +ic^{B\dagger}_{j+1,\sigma}+ic^{A\dagger}_{j,\sigma}+c^{A\dagger}_{j+1,\sigma}
  \right ]|0\rangle. 
\end{equation}
For $U\geq 0$, {\it i.e.}, repulsive interactions, all states are thus
localized as long as the filling is less than half, $\rho\equiv
(N_{\uparrow}+N_{\downarrow})/N_{s}\leq 1/2$, where $N_{\sigma}$ is
the total number of fermions with spin $\sigma$ and $N_{s}=2L$ is the
number of sites. The ground state energy is then trivially given by
$E(\rho\leq 1/2) = -2t N_{s}\rho$, and the chemical potential is
constant, $\mu =-2t$ resulting in infinite compressibility. When
$U<0$, the fermions can lower their energy further by pairing and,
consequently, they are no longer geometrically localized. This is the
situation we shall study here. We note that this model is different
from that treated in Ref.[\onlinecite{sticlet14}], where the spin of
the fermion changes when it performs inter-chain hops but it does not
change for intra-chain hops.

Our system is like that considered in
Ref.[\onlinecite{Tovmasyan2016}], where hopping does not induce spin
change, and where it is shown that the Bardeen-Cooper-Schrieffer (BCS)
wave function is an exact eigenstate of an effective Hamiltonian,
valid at low energies. In this regime, one obtains that the system is
infinitely compressible and the eventual coupling to the upper band
results in a finite compressibility, accompanied by an algebraic decay
of the pair Green function.

\section{Methodology and results}
We employed complementary approaches to study pairing in this
attractive model at different filling fractions. If the attractive
interaction induces up and down fermions to form pairs, the resulting
composite bosons (not necessarily local) may delocalize, yielding a
superfluid/superconducting phase. Such behavior would be signaled by
power law decay of the pair correlation function,
$G^{\alpha\beta}_p(r)$, and exponential decay of the single particle
Green function,
$G^{\alpha\beta}_{\sigma}(r)$~\cite{Giamarchibook,Fradkinbook},

\begin{eqnarray}
\label{pairgreen}
G^{\alpha\beta}_p(r) &=& \langle \Delta^{\alpha\dagger}_{j+r}
\Delta^{\beta}_{j} \rangle,\\ 
\label{singleparticle}
G^{\alpha\beta}_{\sigma}(r) &=& \langle c^{\alpha\dagger}_{j+r,\sigma}
c^{\beta}_{j,\sigma} \rangle,\\
\label{pairoperator}
\Delta^{\alpha}_j &\equiv&
c^{\alpha}_{j,\uparrow}c^{\alpha}_{j,\downarrow},
\end{eqnarray}
where $\Delta^{\alpha}_j$ is a pair annihilation operator at site $j$
on chain $\alpha$, and $\alpha=A,B$ and $\beta=A,B$ label the
chains. We mostly focus on the case where $\alpha=\beta$, {\it i.e.},
intra-chain correlators. As detailed below, we study these correlation
functions by means of DMRG~\cite{White1992,White1993} on large
lattices (up to $L=192$) with open boundary conditions. The remaining
quantities are obtained with periodic boundary conditions, as we
describe below.

Another important physical quantity characterizing transport is the
superfluid weight, $D_s$, given, in one dimension
by~\cite{Kohn1964,Zotos1990,Shastry1990,Fye1991,Scalapino1993,Hayward1995}
\begin{equation}
D_s = \pi L \frac{\partial^2 E_0(\Phi)}{\partial  \Phi^2}\bigg |_{\Phi=0}.
\label{eq:superfluidweight}
\end{equation}
Here, $E_0(\Phi)$ is the ground state energy in the presence of a
phase twist $\Phi$ applied via the replacement $c^{\alpha}_j\to {\rm
  e}^{{\rm i}\phi j} c^{\alpha}_j$, where $\phi =\Phi/L$ is the phase
gradient. This endows the hopping terms with a phase ${\rm exp}({\rm
  i}\phi)$ (or its complex conjugate). \footnote{We emphasize that if
  using open boundary conditions, a local gauge transformation can be
  applied to remove the phases $\phi$ in the hopping terms. Likewise,
  in the case of periodic boundary conditions, a similar
  transformation relates the two equivalent cases: when the phase
  $\Phi$ is spread over the bonds and when they are accumulated at the
  boundaries, such that only one bond has a twist $c^{\alpha}_L\to
  {\rm e}^{{\rm i}\Phi} c^{\alpha}_1$.}
  
As explained in Ref.~[\onlinecite{Scalapino1993}], taking
  the thermodynamic limit, $L\rightarrow\infty$, and computing the
  curvature are non-commuting operations in two and three
  dimensions. The Drude weight, $D$, is obtained by computing the
  curvature for finite lattices, and extrapolating it to the
  thermodynamic limit. On the other hand, the superfluid weight,
  $D_s$, is obtained by computing first the ground state energy
  $E_0(L,\Phi)$ for lattice size $L$, extrapolating to the
  thermodynaimc limit and then calculating the curvature. However, in
  one dimension~\cite{Scalapino1993}, the two operations do commute: the
  curvature of the ground state is essentially related to the Drude
  weight. One, therefore, needs further diagnostics to identify the
  superfluid phase. For bosonic systems, world line algorithms allow
  direct computation of the superfluid
  density~\cite{Rousseau14}. However, for the Creutz lattice, the
  localization of the states due to the flat bands leads to a
  vanishing Drude weight for the non-interacting system and therefore
  the superfluid weight, $D_s$, and the Drude weight, $D$, are
  essentially equivalent and given by
  Eq.~\eqref{eq:superfluidweight}. For the regular two-leg ladder, we
  will see below that this is not so: Even for $|U|\to 0$, the right-hand side of Eq.~\eqref{eq:superfluidweight} does not vanish and actually corresponds to the non-interacting Drude weight. However, this does not mean that the system is superfluid at
  $|U|=0$. To determine if the system is superconducting, we examine
  the single-particle and the pairing Green functions. We identify a 
  superconducting phase by the exponential decay of the former and
  power law decay of the latter. With this caveat in mind, we use the
  notation $D_s$ to which we refer equivalently as the Drude or
  superfluid weight.

\subsection{Superfluid weight - Creutz lattice}
\begin{figure}[t!]
\includegraphics[width=0.99\columnwidth]{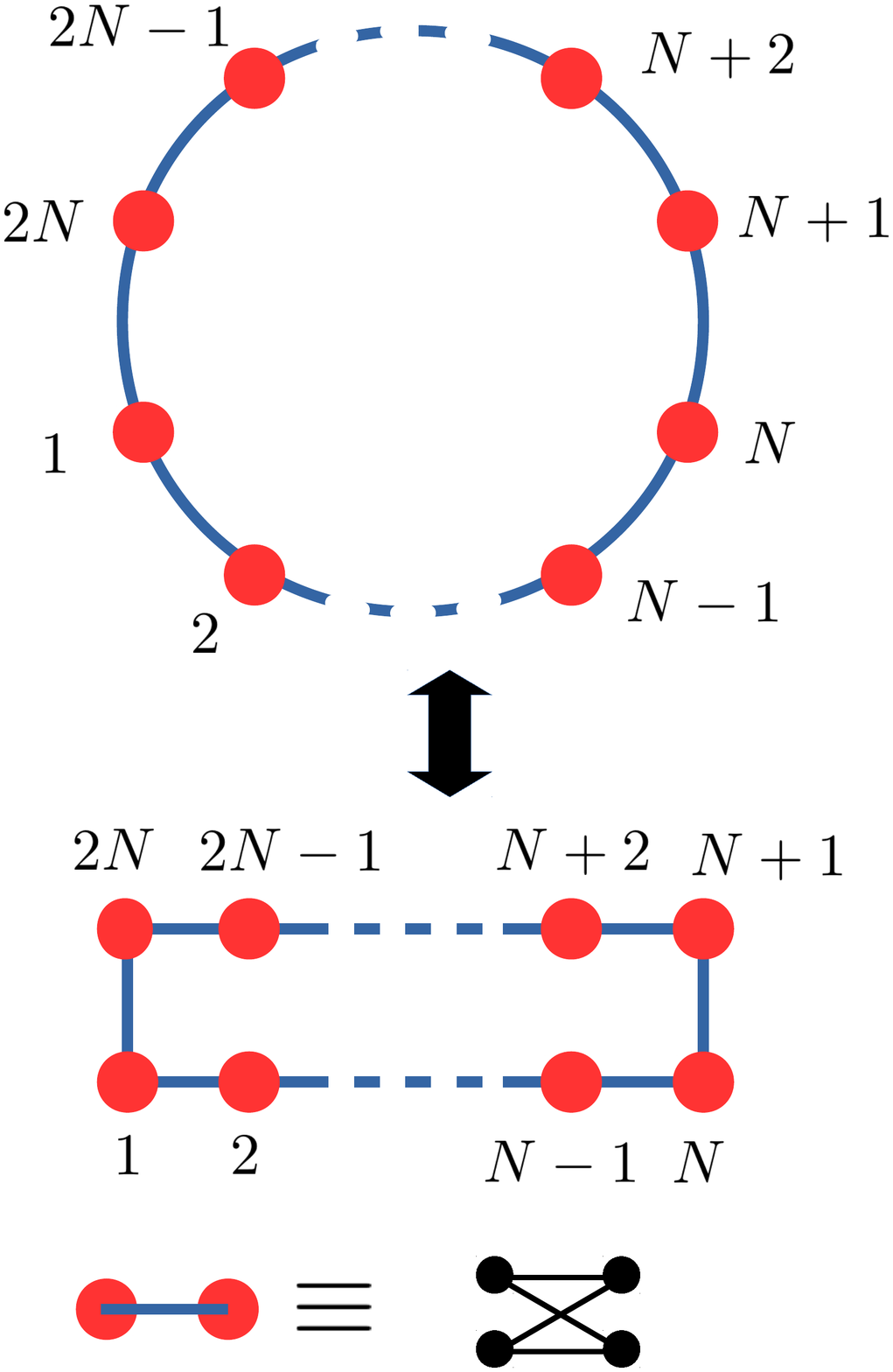}
\caption{(Color online) Mapping a system with periodic boundary
  condition to a ladder-like structure.  The ring shape structure of
  the system with PBC is folded to ladder-like structure with OBC and
  vanishing couplings between the two legs of the ladder for all sites
  but the first and the last.}
 \label{fig:pbc_to_ladder}
\end{figure}

We focused on four different fillings: $\rho_{\rm f} = 1$, $3/4$, $
1/2$ and $1/4$, using a combination of exact diagonalization (ED) for
smaller lattice sizes and DMRG for the larger ones, with periodic
boundary conditions (PBC) in both cases. In the former, we were
restricted to lattices that are commensurate with those fillings, and
such that the reduced Hilbert space sizes, at different momentum
sectors of the translation-invariant Eq~\eqref{eq:creutzham}, are
$\lesssim 10^8$. For example, in the fillings $\rho_{\rm f}=1/2$ and
$1/4$, the largest system sizes tackled using ED had $L=10$ and 16,
respectively. Nonetheless, the DMRG results complemented these for
larger lattices, where we have kept up to $1200$ states in the
truncation process and checked that the results were unchanged when
more states are kept. It is well known that imposing PBC strongly
degrades the efficiency of DMRG. One of the reasons is that PBC
results in an effective long-range coupling between the first and last
sites. On the other hand, as displayed in
Fig.~\ref{fig:pbc_to_ladder}, one can fold the system with PBC to a
ladder-like structure with OBC and vanishing couplings between the two
legs of the ladder for all sites but the first and last two ones. Of
course, this amounts to doubling the size of the local Hilbert space,
{\it i.e.} for each rung, which, in turn, would require a larger bond
dimension. Still, we have checked that for usual 1D chains (bosons or
fermions), the DMRG convergence is much better than with the standard
way of implementing PBC.

\begin{figure}[t!]
\includegraphics[width=0.99\columnwidth]{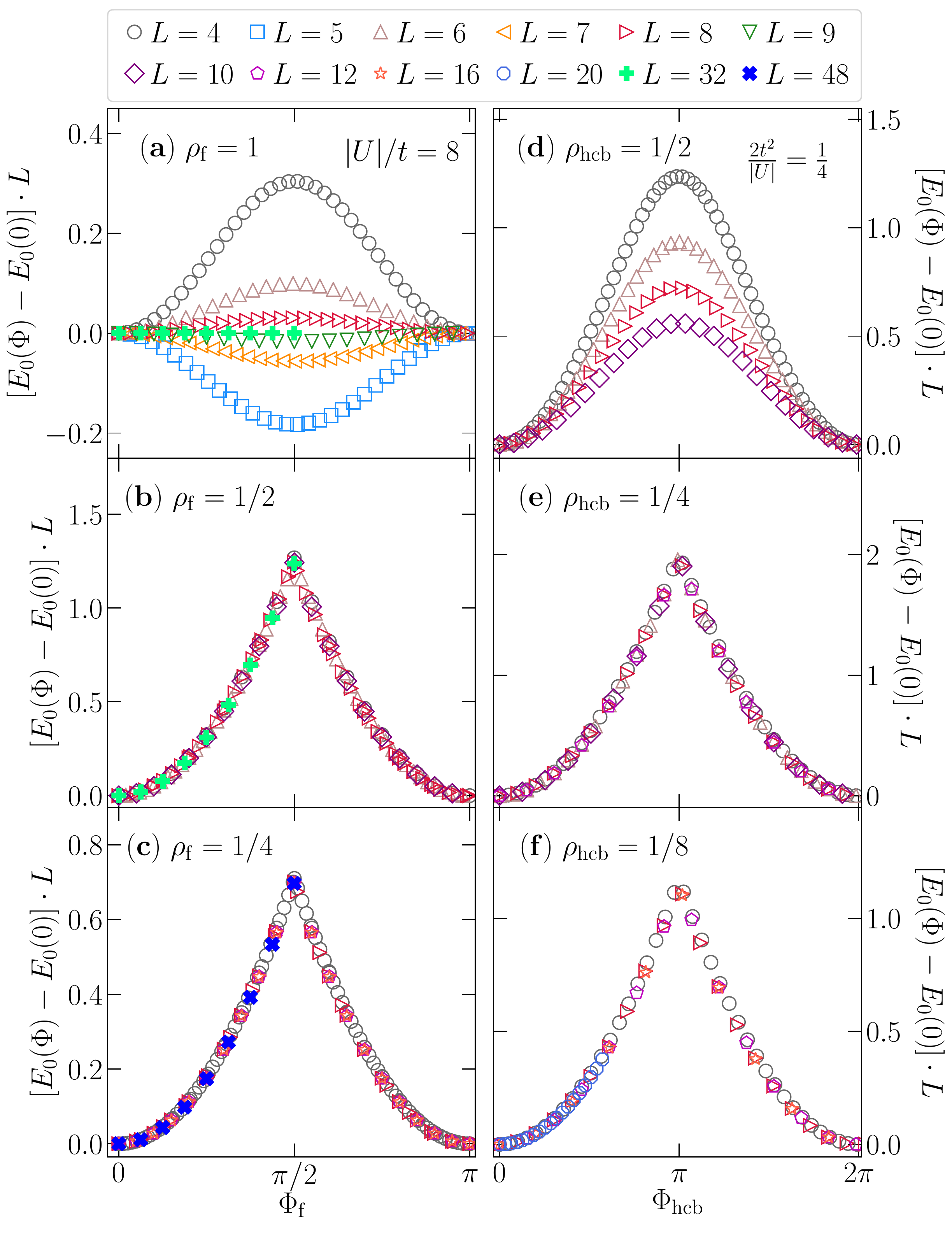}
\caption{(Color online) Dependence of the ground-state energy on the
  applied flux $\Phi$ that threads the lattice, for the fermionic
  Hamiltonian~\eqref{eq:creutzham} [(a)--(c)], and for the effective
  hard-core boson Hamiltonian~\eqref{eq:hardcoreham} [(d)--(f)]. In the
  former, we use $|U|/t=8$ and in the latter, $J=V=2t^2/8$. The
  densities in (a)[(d)], (b)[(e)], and (c)[(f)] are $\rho_{\rm
    f}=1$ [$\rho_{\rm hcb}=1/2$], $\rho_{\rm f}=1/2$ [$\rho_{\rm
      hcb}=1/4$] and $\rho_{\rm f}=1/4$ [$\rho_{\rm hcb}=1/8$]. The
  filled (empty) symbols depict the DMRG (ED) results.}
\label{fig:E0_vs_phi}
\end{figure}

We start by reporting on Fig.~\ref{fig:E0_vs_phi}~[(a)--(c)], the
ground state energy, $E_0$ in Eq.~\eqref{eq:creutzham} with
$|U|/t=8$, as a function of $\Phi_{\rm f}$ [hereafter, $\Phi =
  \Phi_{\rm f}$, for the fermionic Hamiltonian~\eqref{eq:creutzham}],
for different system sizes and fillings. When we subtract the zero
gauge contribution, $E_0(\Phi_{\rm f})=0$, and rescale by the system
size, we notice that, at half filling, the curvature decreases as the
system size increases. This is in stark contrast with the cases with
densities $\rho_{\rm f}<1$ [Figs.~\ref{fig:E0_vs_phi}~(b) and
  ~\ref{fig:E0_vs_phi}(c)], where they are rather insensitive to
$L$. This is the first indication that superfluidity is manifest only
away from half-filling.

\begin{figure}[t]
\includegraphics[width=0.99\columnwidth]{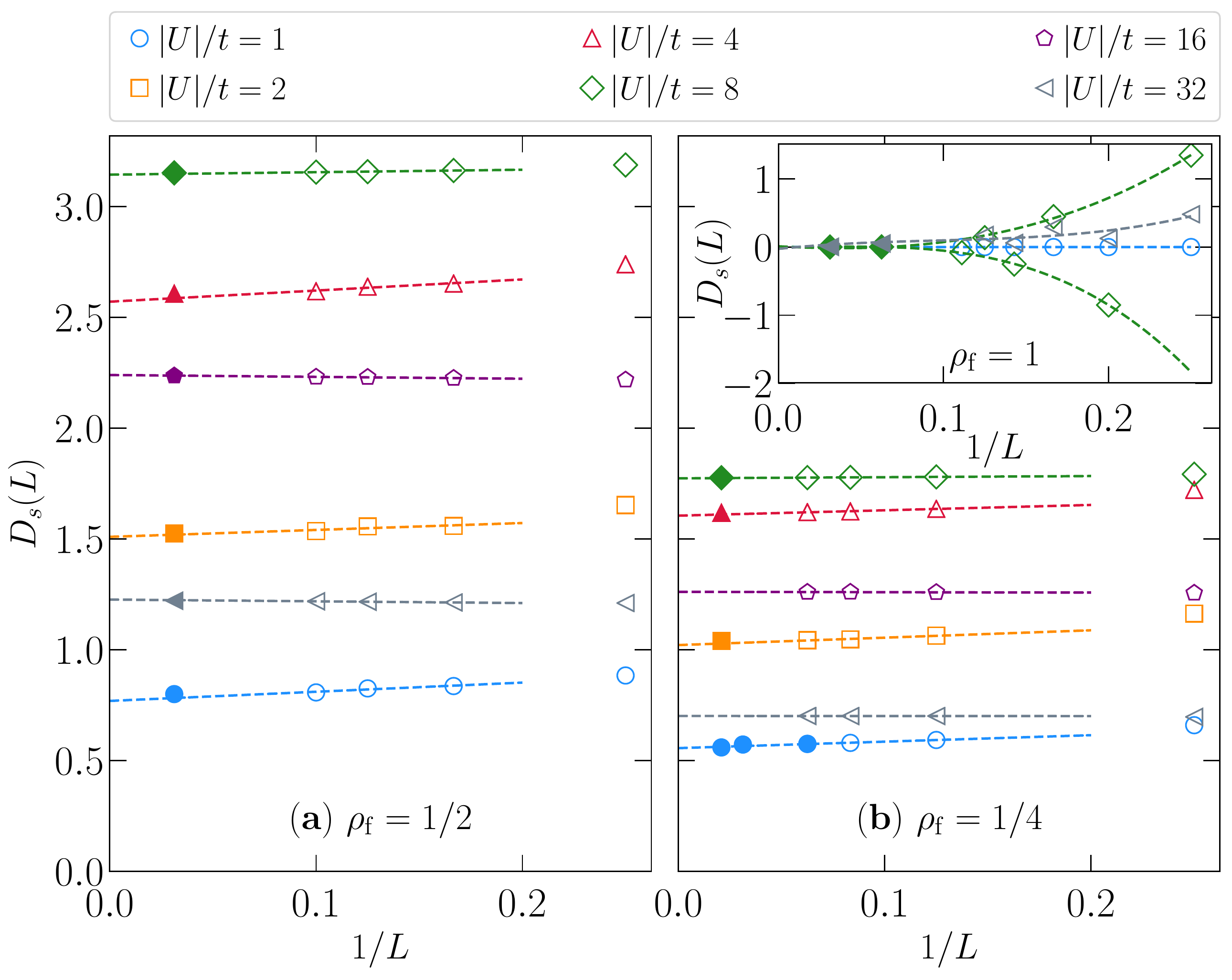}
\caption{(Color online) Extrapolation of $D_s$ to the thermodynamic
  limit for three different fermionic densities, (a) $\rho_{\rm f} = 1/2$, (b) $\rho_{\rm f} = 1/4$ and inset  $\rho_{\rm f} = 1$, with different values of interactions $U$. The open (full) symbols were obtained with ED
  (DMRG); lines in the main panels are linear fits for the larger systems sizes. In the inset, we use an exponential fit and notice a typical even and odd effect related to the formation of closed shells, also seen in other contexts as, for example, in the Drude weight for the 1d Hubbard model.\cite{Fye1991} It is seen that the two methods yield consistent results and give finite extrapolations for $D_s$ when $L\to\infty$, for densities other than $\rho_{\rm
    f}=1$.}
\label{fig:curvature_scaling}
\end{figure}

Finite size scaling of the curvatures is displayed in
Fig.~\ref{fig:curvature_scaling}, to probe the results when
approaching the thermodynamic limit; it shows clearly that
$D_s(L\to\infty)$ is finite for $U<0$ and $\rho_{\rm f}<1$. At
half-filling [inset in Fig.~\ref{fig:curvature_scaling}(b)], the
system displays a vanishing superfluid weight, for a large range of
interactions. Moreover, $D_s$ has a non-monotonic dependence on $U$
away from half-filling. It grows as the strength of the attractive
interactions grows until $|U|/t \approx 8$, where it starts to
decrease in the strongly interacting regime.

\begin{figure}[b!]
\centerline{\includegraphics[width=0.99\columnwidth]{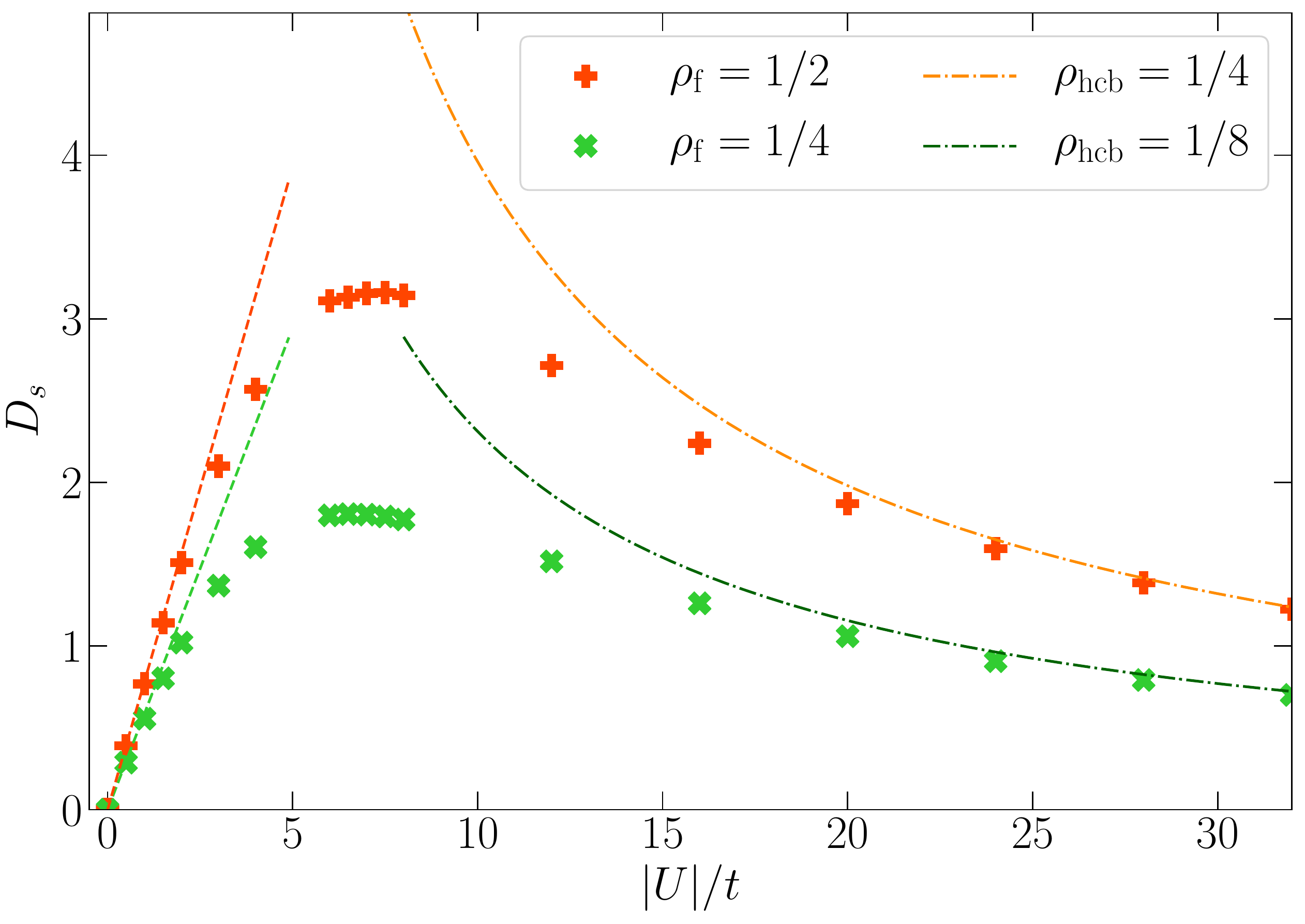}}
\caption{(Color online) $D_s$, for $L\to\infty$, \textit{vs.}
  $|U|/t$, highlighting the non-monotonic behavior of the
  superfluidity. For small interactions, $D_s$ increases linearly with
  slope $\pi\rho(1-\rho)$ [See text]. At large $|U|/t$, the decay can
  be explained by an effective Hamiltonian of repulsive hardcore
  bosons in a Creutz geometry, Eq.(\ref{eq:hardcoreham}). These are
  represented by the dashed-dotted lines.}
\label{fig:Ds_final}
\end{figure}

Finally, by compiling the values of $D_s$ extrapolated to the
thermodynamic limit, we construct in Fig.~\ref{fig:Ds_final} the
dependence of the superfluidity on the interactions for different
densities, where the non-monotonic behavior is evident. Essentially,
the interactions induce a crossover betwen two regimes, at small and
large values of $|U|/t$.  As mentioned in Introduction, and
explained in detail in Ref.~[\onlinecite{peotta15}], the
superfluid weight (Drude weight in 1D) for a multiband system,
computed within a mean-field BCS approach, is the sum of three
different terms. One of the terms is the usual single band BCS term
and vanishes for a flat band, whereas the other two terms have a
topological origin, {\it i.e.} related to the fact that the band
structure has a non-trivial Berry curvature in two dimensions or a
nonzero winding number in one. In the present situation, neglecting
the contribution from the upper band, {\it i.e.} in the limit $U\ll
t$, one has [\onlinecite{Tovmasyan2016}]
  \begin{equation}
  D_s = \pi |U|\rho(1-\rho).
\label{eq:huberDs}
\end{equation}
This linear dependence on $|U|/t$ is plotted in Fig.~\ref{fig:Ds_final}
for $\rho=1/2, \,1/4$, as dashed lines, and is seen to be in excellent agreement with
our exact numerical values for small $|U|/t$.

On the other hand, at large $|U|/t$, the $\uparrow$ and $\downarrow$
fermions form strongly bound pairs being approximately described by a
\textit{local} bosonic particle. Given the constraints on the
occupancy for each site, one can show that in this case, the fermionic
Hamiltonian, Eq.\eqref{eq:creutzham}, can then be mapped onto a
Hamiltonian of \textit{repulsive} hardcore bosons whose hopping and
near-neighbor repulsion have the same energy scale~\cite{Micnas1990},
and the density of particles is $\rho_{\rm hcb} = \rho_{\rm f}/2$. The
effective Hamiltonian then reads

\begin{eqnarray}
{\cal H}_{\rm eff} = -\frac{2t^2}{|U|}\sum_{j,\alpha,\beta} \bigg (
&&b^{\alpha\dagger}_jb^{\beta}_{j+1} + {\rm H.c.}\bigg ) \nonumber
\\ +\frac{2t^2}{|U|}\sum_{j,\alpha,\beta} \bigg ( && n^\alpha_j
n^\beta_{j+1} + n^\beta_{j+1}n^\alpha_j \bigg ),
\label{eq:hardcoreham}
\end{eqnarray}
where $n^\alpha_j = b^{\alpha\dagger}_jb^\alpha_j$, and
$b^{\alpha\dagger}_j$ ($b^\alpha_j$) is a hardcore boson creation
(annihilation) operator on site $j$ and chain $\alpha=A,\,B$. They
satisfy $\{b^\alpha_j,b^{\alpha\dagger}_j\}=1$, and
$[b^\alpha_j,b^{\beta\dagger}_r]=0$ for $j\neq r$ or
$\alpha\neq\beta$. We note that this effective model is defined on the
Creutz lattice, Fig.~\ref{fig:creutzlattice}, but is not governed by a
Creutz Hamiltonian as in Eq.~\eqref{eq:creutzham}, \textit{i.e.} the hopping
bonds are preserved but in this case they all have the \textit{same}
hopping energy. Thus, the Hamiltonian Eq.~\eqref{eq:hardcoreham} is
not flat: It describes hardcore bosons on a quasi-one dimensional
lattice with a non-flat dispersion relation and which are expected to
be superfluid.

Analogously to what we have done with the fermionic Hamiltonian, we numerically
studied the effect of a phase gradient on the hopping terms,
implemented via the replacement $b^{\alpha}_j\to {\rm e}^{{\rm i}\phi
  j} b^{\alpha}_j$ [$\phi/L\equiv \Phi_{\rm hcb}$], in order to probe
the superfluid properties of this effective model. The dependence of
the ground state energy on the flux $\Phi_{\rm hcb}$ is shown in
Fig.~\ref{fig:E0_vs_phi}[(d)--(f)], using ED for different system
sizes. First, we notice the curvatures display a similar qualitative
behavior: At half-filling, the curvature of $E_0(\Phi)\cdot L$
decreases for increasing system sizes while it is independent of $L$
for densities away from it. Second, the periodicity of the $E_0(\Phi)$
curve in the bosonic case is twice as large as in the fermionic
one. This is an expected result based on considerations of flux
quantization of superconducting rings~\cite{ByersYang1961,Yang1962}. A
magnetic flux enclosed by such a ring is related to a vector potential
manifested along the direction of the lattice as $\vec A =
(\varphi/L)\hat x$. This, in turn, results in a twist of the boundary
conditions along the same direction of the form $\exp\{{\rm
  i}2\pi\varphi/\varphi_0\}$, where $\varphi_0$ is the flux quantum
$hc/e$. Byers and Yang~[\onlinecite{ByersYang1961}] have shown that
the energies are periodic functions of $\varphi$, whose period is
$\varphi_0/n$, where $n$ is the total charge of the basic group. For
instance, for superconductors with carriers of charge $2e$, $n=2$. In
our units, the period $\Phi$ of the ground state energy in the
fermionic problem is $\pi$, which correcting the $2\pi$ factor,
results in periods $\varphi=\varphi_0/2$, {\it i.e.} Cooper pairing
superconductivity, as expected. On the other hand, for the case of
effective hardcore boson model, the period of the ground state energy
with $\Phi$ is $2\pi$, that results in periods $\varphi=\varphi_0/1$,
{\it i.e.} the charge of the superfluid carrier is one, a hardcore
boson itself.

Beyond the qualitative description of the mapping between the two
models, we show that the agreement is also quantitative for large
interactions. Computing the superfluid weight via
Eq.~\eqref{eq:superfluidweight} and extrapolating to the thermodynamic
limit (not shown), we obtain $D_s$ for the hardcore boson effective
model [Eq.~\eqref{eq:hardcoreham}] which we display as dashed-dotted
lines in Fig.~\ref{fig:Ds_final}. For large values of $|U|/t$, the
\textit{fermionic} superfluid weight asymptotically approaches the one
for hardcore bosons, further confirming the local nature of the pairs
in this regime and its superfluid character for densities away from
half-filling.

\begin{figure}[t!]
\centerline{\includegraphics[width=0.99\columnwidth]{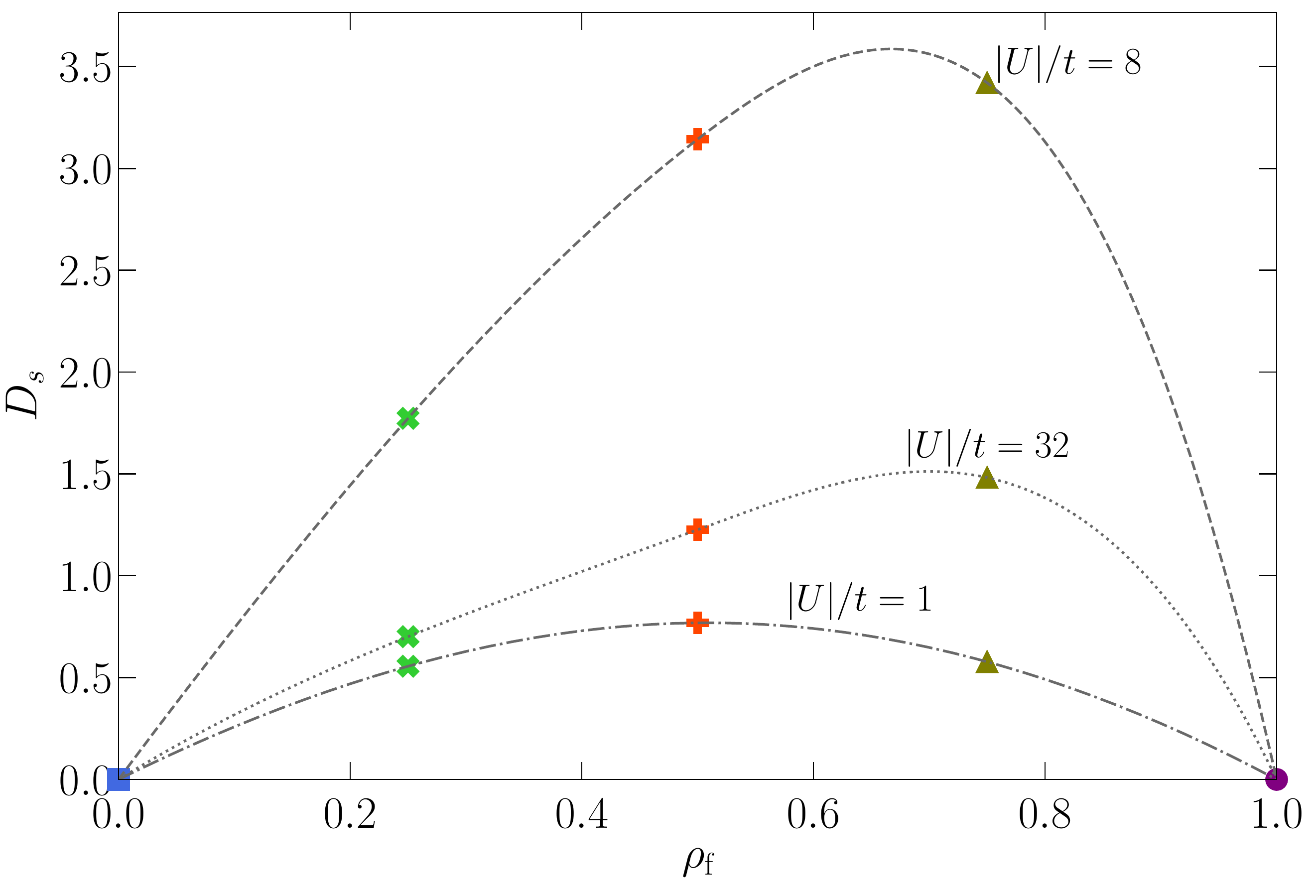}}
\caption{(Color online) Dependence of the superfluid weight on the
  total fermionic density, for different interaction strengths: With
  small interactions ($|U|/t=1$), where the description of the
  projected Hamiltonian is suitable, at large interactions
  ($|U|/t=32$) and around the peak of $D_s$ for the intermediate
  fillings ($|U|/t=8$).}
\label{fig:Ds_vs_rho}
\end{figure}

Turning back to the fermionic problem, Fig.~\ref{fig:Ds_vs_rho}
displays the dependence of the superfluid weight on the total density
$\rho_{\rm f}$ for different interaction strengths.  In the
noninteracting case, due to the dispersionless nature of the bands,
the superfluid weight is zero regardless of the density
investigated. When the interactions are finite but small, such that
the ground state can still be described by a BCS wavefunction, the
superfluid weight follows a form predicted by Eq.~\eqref{eq:huberDs},
symmetric in the densities around $\rho_{\rm f}=1/2$. Away from this
regime, increasing the interactions, causes $D_s$ to become
asymmetric, with its peak moving to higher filling.

\subsection{Excitation gaps}

\begin{figure}[t!]
\includegraphics[width=0.99\columnwidth]{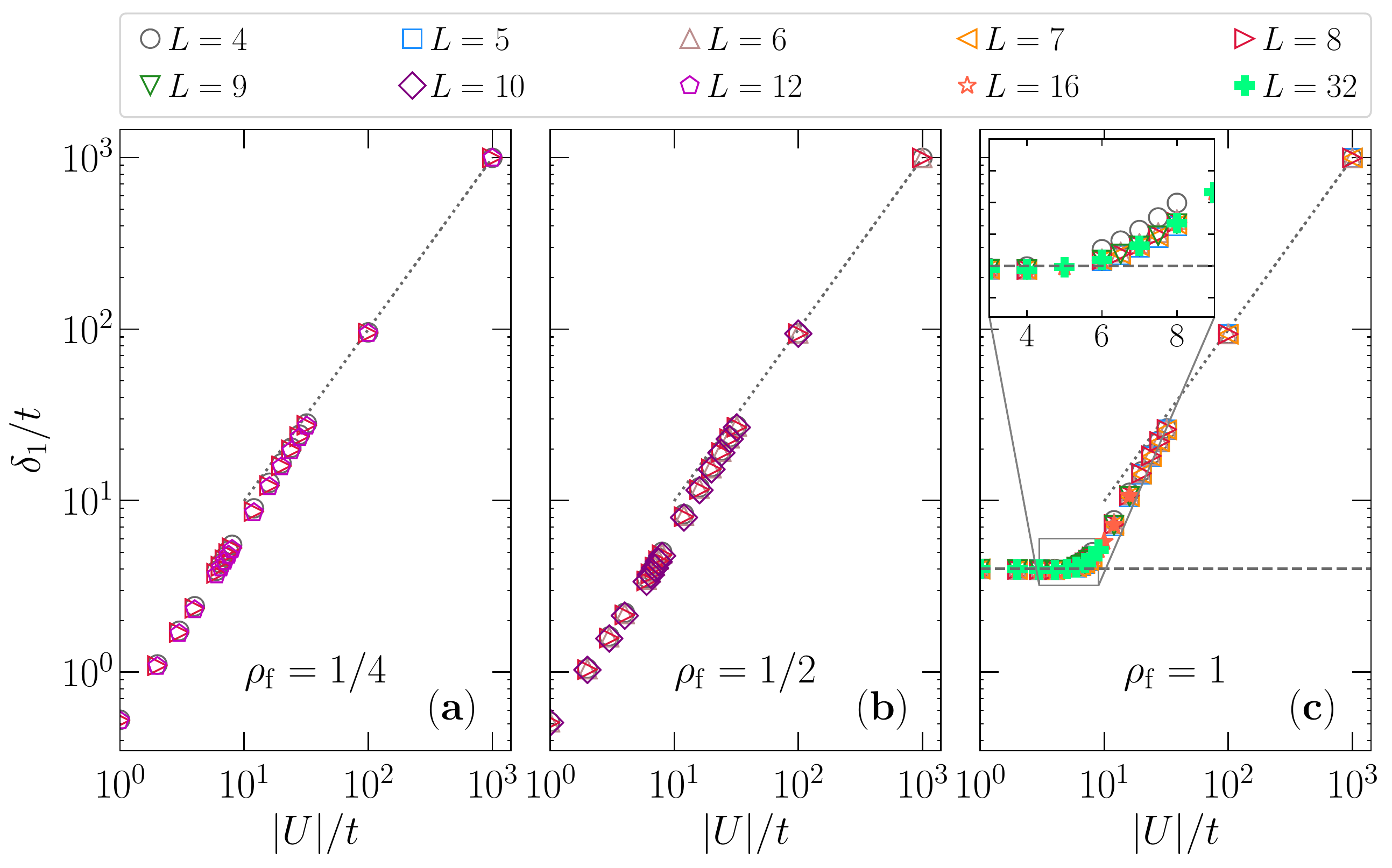}
\caption{(Color online) The one-particle gap energies for
  $\rho=1/4,\,1/2$ and $1$ as functions of the coupling $|U|/t$. The
  dotted lines have the form $\delta_1\propto |U|/t$ and are guides to
  the eye. For $\rho=1$ [(c)], the dashed line at $\delta_1=4t$
  indicates the range of validity of the projected Hamiltonian, where
  excitations are gapped by the non-interacting Creutz bandwidth. The
  inset displays in detail the departure from this regime, starting at
  interactions $|U|/t\gtrsim 4$. As in previous figures, ED (DMRG)
  results are given by empty (filled) symbols.}
\label{fig:single_particle_gap}
\end{figure}

We further characterize the transport properties of the system by
studying the nature of particle excitations. In particular, we study
the fate of one- and two-particle excitations on the Creutz Hubbard
Hamiltonian to understand better the superfluid behavior. The
$m$-particle excitation energy, {\it i.e.} the energy gap per particle
to promote such excitation, can be defined
as~\cite{Dodaro2017,Junemann2017}
\begin{equation}
\delta_m \equiv \frac{1}{m}\left [ E_0(N+m) + E_0(N-m) -2E_0(N) \right ],
\label{eq:energygaps}
\end{equation}
where $m$ is the number of doped particles in a system with $N$
particles; $E_0$ is the corresponding ground state at those
fillings. We first describe the single particle ($m=1$) excitations in
Fig.~\ref{fig:single_particle_gap}, for the densities $\rho_{\rm
  f}=1,\,1/2,$ and $1/4$, as functions of $|U|/t$. They show that the
gap to add one particle has small finite size corrections;
furthermore, in the regime of strong interactions, the gaps are
proportional to $|U|/t$, indicating tighter binding. At small
interaction strengths, the behavior of $\delta_1$ is markedly
different for different densities. While for $\rho_{\rm f}<1$, the
single-particle gap is finite and proportional to $|U|/t$, at
half-filling, $\delta_1 \gtrsim 4t$, suggesting that the
single-particle picture, with two flat bands separated by a gap of
this same energy, is still applicable. For that density, the lower
band is completely filled and the cost in energy to add an extra
particle is then $4t$.

\begin{figure}[t!]
\includegraphics[width=0.99\columnwidth]{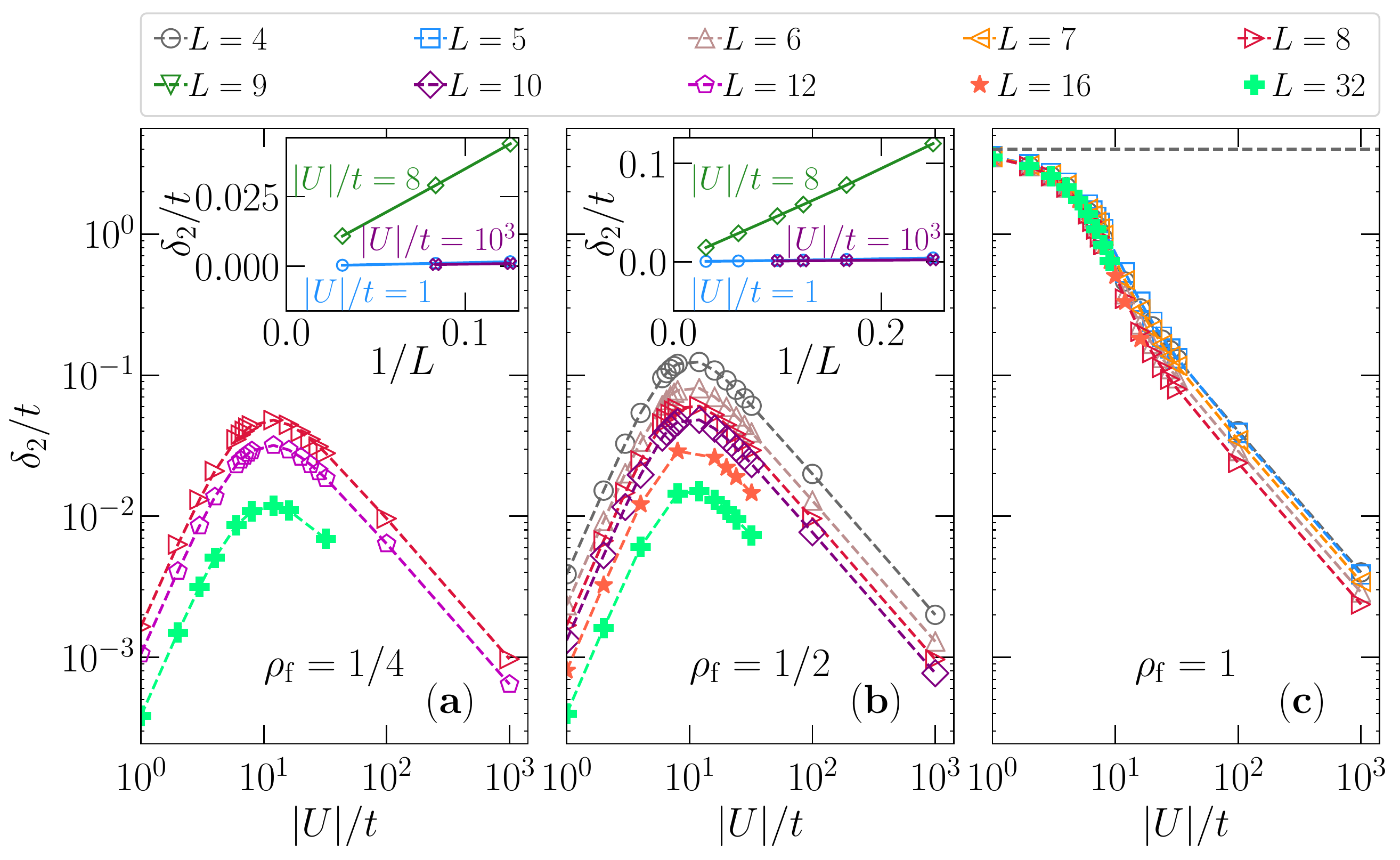}
\caption{(Color online) Similar to Fig.~\ref{fig:single_particle_gap},
  but for two-particle excitations. The dashed line at $\delta_2=4t$
  in (c) indicates the range of validity of the projected
  Hamiltonian. As in previous figures, ED (DMRG) results are given by
  empty (filled) symbol. The insets in panel (a) and (b) display the
  finite size scaling of $\delta_2$ for three values of the
  interactions at densities $\rho_{\rm f} = 1/4$ and $1/2$,
  respectively.}
\label{fig:two_particle_gap}
\end{figure}

Figure~\ref{fig:two_particle_gap} shows that the two-particle gaps
suffer appreciable finite size effects. Away from half-filling, the
dependence of $\delta_2$ on $|U|/t$ suggests that this quantity peaks
at interactions corresponding to the maximum of the superfluid weight,
$|U|/t\approx8$. However, finite size scaling analysis [insets in
  Fig.~\ref{fig:two_particle_gap}(a) and
  ~\ref{fig:two_particle_gap}(b)] shows that this pair excitation is
gapless, as one would expect for a system displaying finite superfluid
weight in the thermodynamic limit. For $\rho_{\rm f}=1$ and small
interactions, the energy per particle to add a pair is again close to
the noninteracting gap. It takes values slightly below $4t$ in this
regime because the added particles, which populate the upper band, can
further decrease their energy by interacting attractively. Most
importantly, due to the minimal dependence on system size, one can
guarantee that the system is not superfluid, in agreement with the
results of Fig.~\ref{fig:curvature_scaling} at this filling. In the
strongly interacting regime, finite size effects are more pronounced,
and $\delta_2$ steadily decreases for larger $L$'s. Nonetheless, we
can once again resort to the mapping to the effective Hamiltonian~\eqref{eq:hardcoreham} in this regime, to settle the question
whether the two-particle gaps are finite.

\begin{figure}[h!]
\includegraphics[width=0.99\columnwidth]{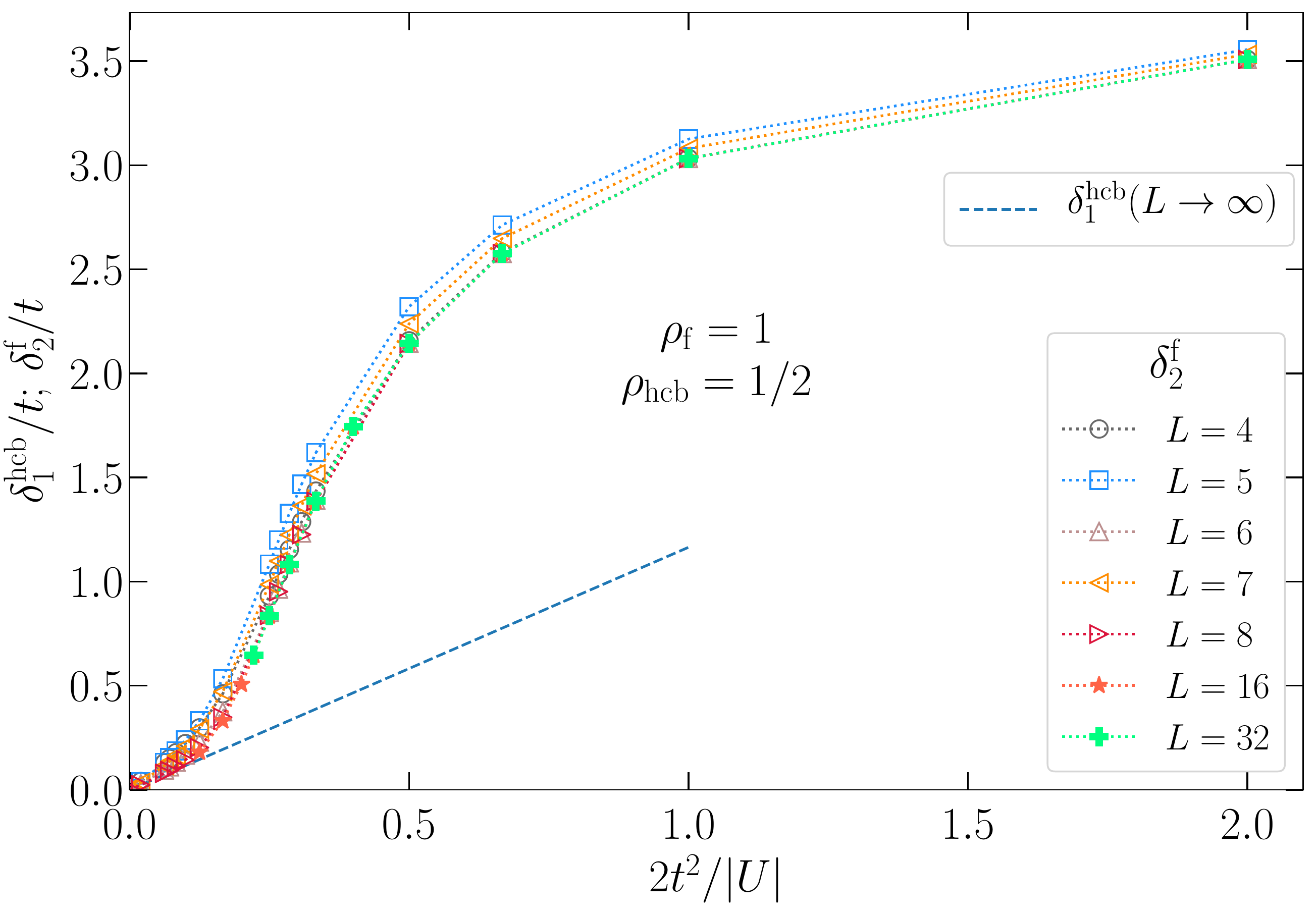}
\caption{(Color online) Single-particle gap for the effective hard
  core boson model~\eqref{eq:hardcoreham}, $\delta_1^{\rm hcb}$,
  and two-particle gap for the original fermionic model~\eqref{eq:creutzham}, $\delta_2^{\rm f}$, as functions of
  $2t^2/|U|$. The extrapolation to the thermodynamic limit of
  $\delta_1^{\rm hcb}$ is given by the dashed line, whereas
  $\delta_2^{\rm f}$ is presented for different system sizes. As in
  previous figures, ED (DMRG) results are given by empty (filled)
  symbols.}
\label{fig:single_particle_gap_hc_bosons_and_two_particle_gap_fermions}
\end{figure}

%
In the coupling regime where the pairs are tightly bound, one expects
the two-particle fermionic gap [$\delta_2^{\rm f}\equiv\delta_2$] to
correspond to the single particle gap of hardcore bosons,
$\delta_1^{\rm hcb}$, in the effective model language. By performing
finite size scaling on the latter, we find that $\delta_1^{\rm hcb}$
is finite in the thermodynamic limit as long as the interactions are
also finite. Consequently, one expects $\delta_2^{\rm f}$ to be finite
too at strong interactions.
Figure~\ref{fig:single_particle_gap_hc_bosons_and_two_particle_gap_fermions}
displays the dependence of $\delta_2^{\rm f}$ and $\delta_1^{\rm
  hcb}$, at half-filling, as functions of the inverse interaction
strength. The agreement of the single particle gap for hardcore
bosons and the two-particle gap for fermions is evident at $|U|\gg
t$. Essentially, $\delta_1^{\rm hcb}$ provides a lower bound to
$\delta_2^{\rm f}$, indicating the gap to create a pair is always
finite as long as $|U|$ is. This again confirms the picture that the
superfluid weight for $\rho_{\rm f}=1$ is zero at arbitrary values of
the interactions in the Creutz ladder.

\subsection{Pair correlation functions}

\begin{figure}[h!]
\includegraphics[width=0.99\columnwidth]{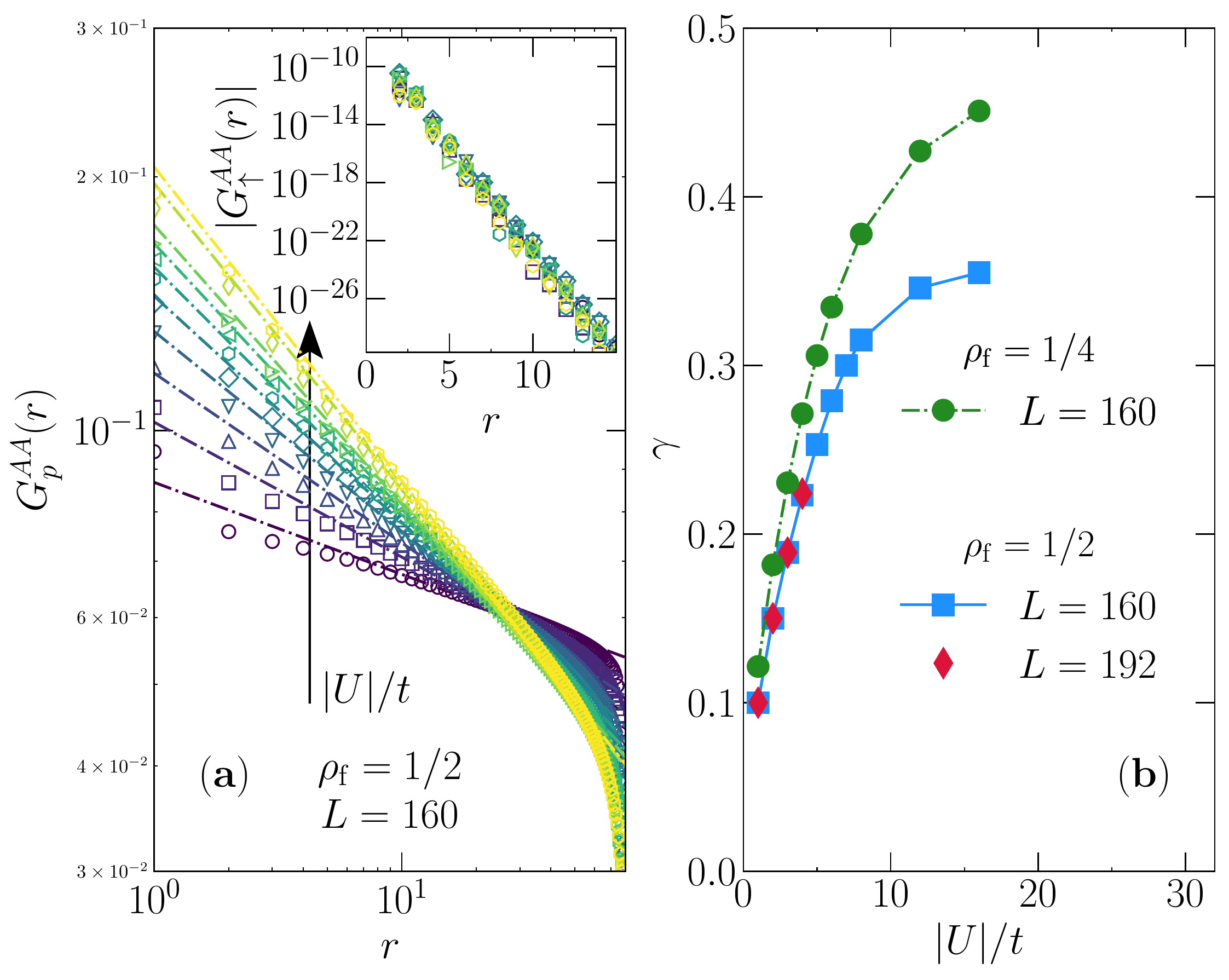}
\caption{(Color online) (a) The intra-chain pair correlation function,
  $G_p^{AA}$ as a function of the distance in a log-log scale and
  different values of $|U|/t = 1, 2, 3, 4, 5, 6, 7, 8, 12, 16$, as
  signaled by the vertical arrow. The lattice size is $L=160$ and
  $\rho_{\rm f} = 1/2$. The inset displays the single-particle Green's
  functions in a linear-logarithm scale. In (b), the dependence on $|U|/t$
  of the decay exponent $\gamma$ of the pair Green function, for
  densities $\rho_{\rm f} = 1/4$ and $1/2$. At large attractive
  interactions, it possesses an asymptotic behavior, saturating at the
  value corresponding to the decay of single-particle correlations,
  $\langle b^{\dagger}_{j,r}b_{j}\rangle$, of repulsive hardcore
  bosons in the same geometry and density $\rho_{\rm hcb} = \rho_{\rm
    f}/2$. Finite size effects are already rather small for the values
  of $L$ considered, 160 and 192, at density $\rho_{\rm f} = 1/2$.}
\label{fig:Gpair}
\end{figure}

The finite gaps for the single particle excitation suggest that the
single particle Green's function, Eq.~\eqref{singleparticle}, should
decay exponentially. On the other hand, the high pair mobility
indicated by the vanishing of the two-particle excitation energy
suggests that the pair Green function, Eq.~\eqref{pairgreen}, decays
as a power with distance, for densities away from half-filling. This
is confirmed in the inset of Fig.~\ref{fig:Gpair}(a), which shows very
fast exponential decay of the one-particle Green function with
distance, signaling a robust single-particle gap at the density
$\rho_{\rm f}=1/2$. We have also confirmed that similar behavior occurs
for other densities. Moreover, we remark that $\uparrow$ or
$\downarrow$ channels, in either chain $A$ or $B$, result in
equivalent values for this correlation.

In contrast, the power-law decay of the pair Green function
[Fig.~\ref{fig:Gpair}(a)] is characteristic of quasi-long range order
for this observable, and indicates that local pair excitations are
gapless in this system. We note that the larger the attractive
interaction the faster is the decay of $G_p^{\alpha\alpha}$ (we use
$\alpha=A$). When compiling the values of the decay exponent $\gamma$
in Fig.~\ref{fig:Gpair}(b), where $G_p^{\alpha\alpha}\propto
r^{-\gamma}$, we note that it essentially saturates at large
interactions, denoting that the extent of the decay of the
correlations is constant. Once more, we can understand this result via
the mapping onto the effective repulsive hardcore boson model in the
Creutz geometry Eq.\eqref{eq:hardcoreham}. In this case, changing the
magnitude of $U$ accounts only for a re-scaling of the Hamiltonian
energies, since both hopping and nearest-neighbor interactions have
the same energy dependence on $|U|/t$, without changing the decay
extent of the correlation functions. Thus, one would expect that
$\langle \Delta^{\alpha\dagger}_{j+r} \Delta^{\alpha}_{j} \rangle
\propto \langle b^{\alpha\dagger}_{j+r}b^{\alpha}_{j}\rangle\propto
r^{-\gamma}$, in the large $|U|/t$ limit.

In (quasi-)one dimensional systems, repulsive hardcore bosons behave
as Luttinger liquids when away from the half-filling
regime~\cite{Crepin2011}. We then expect the exponent $\gamma$ we
obtain in the strongly interacting regime to be related to the
Luttinger liquid parameter $K$, which is a function of the density of
particles. To the best of our knowledge, this is not known for a
system where the interactions and hoppings have the geometry of the
Creutz lattice, thus a quantitative prediction is hard to
make. Nevertheless, the behavior of $D_s$, $\delta_1$, $\delta_2$, and
the Green functions leads to the conclusion that the attractive
Hubbard model in the Creutz lattice, given by
Eq.~\eqref{eq:creutzham}, exhibits superfluidity for any $U<0$, for
densities away from half-filling.

\subsection{Regular two-leg ladder}

\begin{figure}[t]
\centerline{\includegraphics[width=0.99\columnwidth]{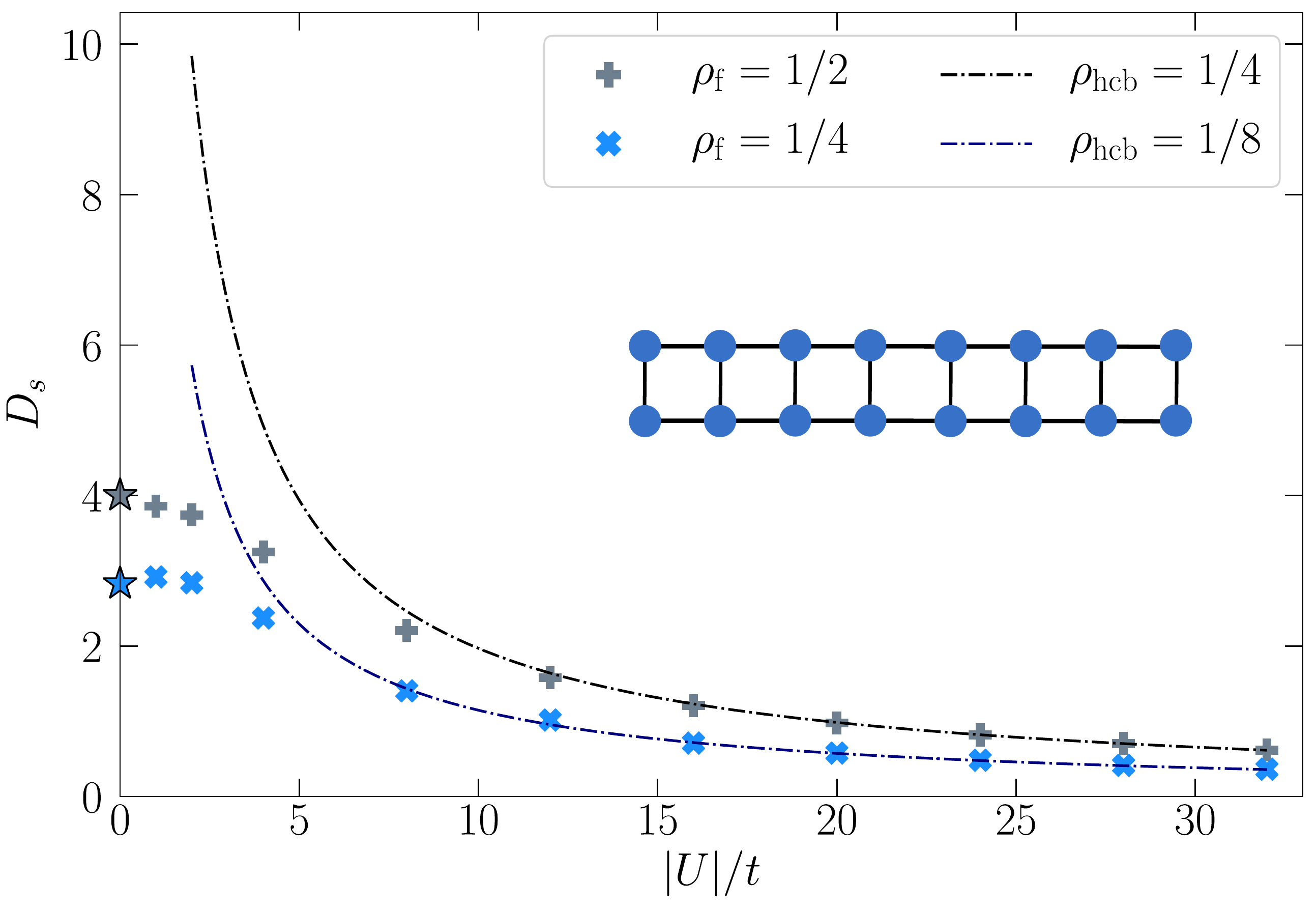}}
\caption{(Color online) Drude weight in the thermodynamic limit
  \textit{vs.} $|U|/t$, for a regular ladder. The non-interacting
  results, which are finite, unlike in the Creutz lattice, are
  indicated by the star symbols at $|U|/t=0$. The strongly interacting
  regime is explained by the results of repulsive hardcore bosons on
  a regular ladder, depicted by the dashed-dotted lines.}
\label{fig:Ds_ladder}
\end{figure}

We now compare the above results of the Creutz lattice with the
behavior of the attractive fermionic Hubbard model on a simple regular
ladder composed of two coupled chains (See
Fig.~\ref{fig:Ds_ladder}). In this case, the Hamiltonian is written
as,
\begin{equation}
{\cal H} = -t \sum_{\langle i,j\rangle,\sigma} \left (
c^{\dagger}_{i,\sigma} c^{\phantom\dagger}_{j,\sigma} + {\rm H.c.}\right ) +
U\sum_j n_{j,\uparrow} n_{j,\downarrow},
\label{eq:ladderham}
\end{equation}
where $\langle i,j\rangle$ indicates inter- and intra-chain nearest
neighbors and, again, $U<0$. To start, we highlight the main
difference between the Creutz and regular ladders, which is evident
when comparing the non-interacting regime: In the Creutz case, as
$|U|/t\to 0$, $D_s\to 0$ due to the geometrical localization caused by
the flat band; on the other hand, in the ladder case, $|U|\to 0$ leads
to a free fermion gas with non-flat dispersion which is mobile.

For finite interactions, we resort to numerical calculations. We
calculate the superfluid weight via Eq.~\eqref{eq:superfluidweight},
by repeating the analysis done for the Creutz lattice. After finite
size extrapolations to the thermodynamic limit, we obtain the
dependence of $D_s$ on the interaction strength depicted in
Fig.~\ref{fig:Ds_ladder}. As before, one can explain the strongly
interacting limit using an effective hardcore boson Hamiltonian
similar to Eq.~(\ref{eq:hardcoreham}), but with hopping and
interacting terms corresponding to the ladder geometry. However, in
contrast to the Creutz case, the non-interacting, $U=0$, result is
finite and can be calculated exactly using the energy dispersion of
the ladder, $\varepsilon_{k} = -2t\cos(k) \pm t$. The definition of
the superfluid weight [Eq. (\ref{eq:superfluidweight})] yields
$D_s(U=0) = 4t\sin\left(\pi\rho\right)$, for $\rho \leq1/2$, {\it
  i.e.}  where only the lower band has finite occupancy in the ground
state. These are indicated by star symbols in Fig.~\ref{fig:Ds_ladder}
for the two densities we investigated, $\rho_{\rm f} = 1/4$ and
$1/2$. However, it is very important to keep in mind our
  discussion after Eq.(\ref{eq:superfluidweight}): That $D_s(U=0)$ is
  nonzero does not mean that the noninteracting system is
  superfluid. At $U=0$, both the single particle and pair Green
  functions decay as power laws indicating metallic behavior. Our
  results indicate that (up to the precision of our calculations) as
  soon as $|U|\neq 0$, the fermions start to pair and form a
  superfluid phase characterized by exponential decay of the single
  particle Green function and power law for the pair correlations.

At half-filling, one can apply a particle-hole transformation in one
of the spin components, say $\tilde c_{i,\downarrow}=(-1)^i
c_{i,\downarrow}^\dagger$, keeping the other component unchanged, to
map the Hamiltonian onto the repulsive Hubbard model in a two-leg
ladder. In this case, one expects a Mott insulating behavior, whose
charge stiffness approaches zero in the thermodynamic limit. Hence, in
the original Hamiltonian, both the superfluid and 
  Drude weights should also decay to zero when $L\to\infty$, provided
the interactions are finite.

\begin{figure}[h!]
\includegraphics[width=0.99\columnwidth]{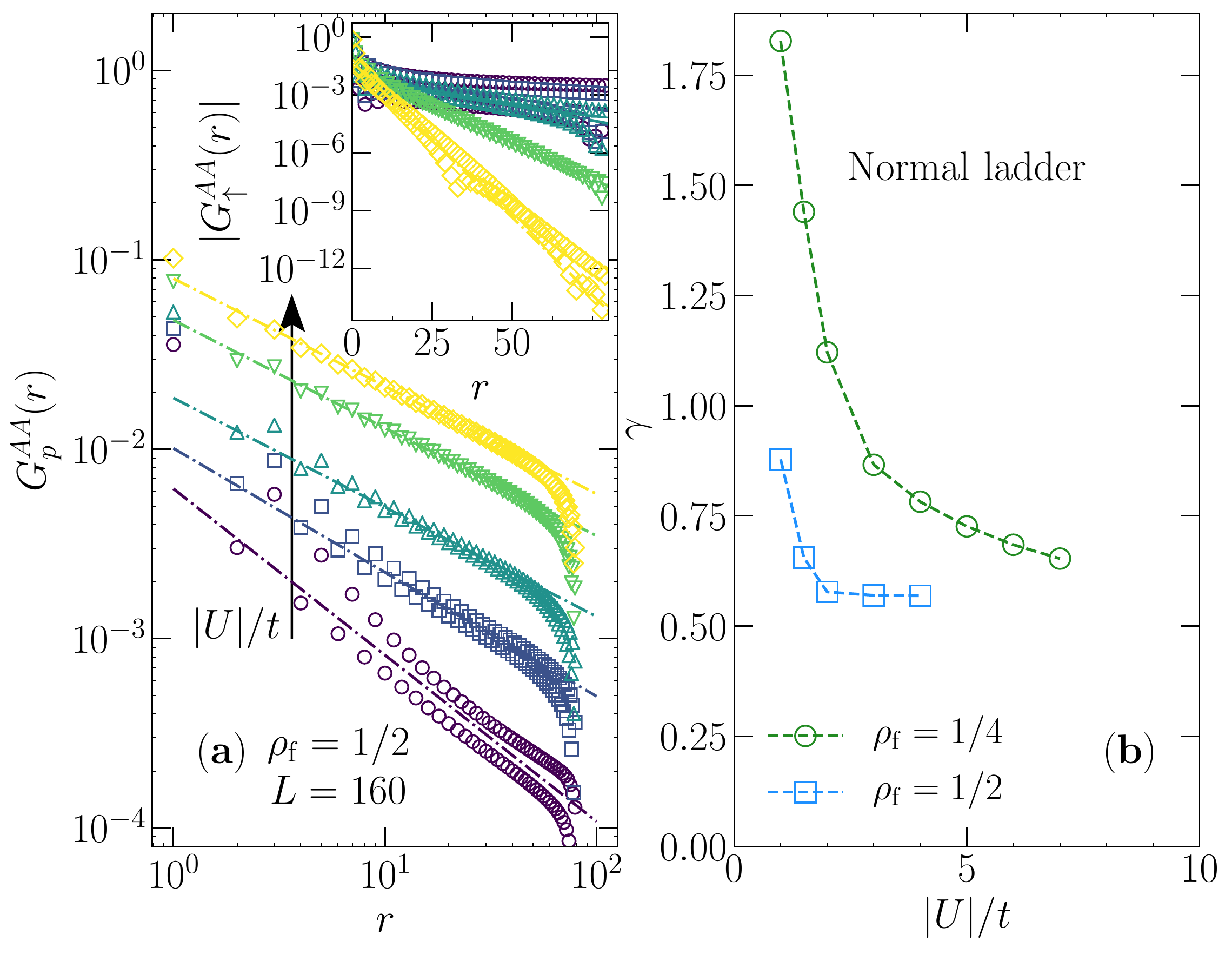}
\caption{(Color online) The same as in Fig.~\ref{fig:Gpair}, but now
  in the case of a regular ladder. The interaction values used are
  $|U|/t = 1, 1.5, 2, 3$ and $4$, and are schematically represented by
  the vertical arrow.}
\label{fig:Gpair_ladder}
\end{figure}

Previous studies using zero-temperature quantum Monte Carlo
techniques~\cite{Guerrero2000} obtained the power-law decay of pair
correlations for the normal ladder at $\rho_{\rm f}=1/2$. The decay
exponent was found to be either $\gamma=1.07(3)$ or $\gamma=0.87(2)$,
if considering the fitting to the upper or lower envelope of the
oscillating pair correlations with distance, for interactions
$|U|/t=2$. Here, we focus on the same density using DMRG. Similarly to
Fig.~\ref{fig:Gpair}, we report in Fig.~\ref{fig:Gpair_ladder}(a) the
decay with distance of the pair and single-particle Green's
functions. Again, the respective power-law and exponential decays
signal the superfluid character of the system and agrees with the
predictions of the superfluid weight presented in
Fig.~\ref{fig:Ds_ladder}. Furthermore, we show in
Fig.~\ref{fig:Gpair_ladder}(b) the dependence of the decay exponent of
the pair Green's functions on the interaction strength. One can
highlight two points: The first is that the magnitude of the
interactions that yields a saturated exponent is much smaller than in
the case of the Creutz lattice (note the different ranges of interactions in both plots). This can be understood by noticing
that the agreement of the superfluid weight of hardcore bosons with
the one obtained for the original fermionic model
(Fig.~\ref{fig:Ds_ladder}) appears at smaller interactions in
comparison to the Creutz lattice (Fig.~\ref{fig:Ds_final}). The second point 
is that at the same density we have investigated for the Creutz ladder
[Fig.~\ref{fig:Gpair}(b)] the decay exponent is larger for the
ladder. This means that the pair correlations, although still
displaying algebraic behavior, are shorter ranged for the ladder than
for the Creutz lattice at the same density. In that sense, one can argue that the
superfluid nature in a Creutz lattice is more robust that in a regular
ladder.

\section{Conclusions and Remarks}

We investigated the superfluid properties of attractive fermions on a
cross-linked ladder using numerical unbiased methods, namely exact
diagonalization and density matrix renormalization group. The ladder,
known as the Creutz lattice, is constructed in such a away as to
render two flat bands in the tight-binding regime. We introduce local
attractive interactions between fermions and show that the system
displays finite superfluid weight with two distinct regimes, of weak
and strong interactions if the fermionic filling is smaller than
one. In the former, we show that this quantity is explained by the
analysis of a projected Hamiltonian on the lower band, valid at small
energy scales~\cite{Tovmasyan2016}, whereas in the latter, it can
described by the superfluid properties of repulsive hardcore bosons on
a similar lattice, but with all the bonds having hopping energies with
the same sign.

Quantitative study of single and two-particle excitations of the
fermionic problem corroborates this picture, showing that the energy
to excite a single charge is gapped for the wide range of interactions
we investigate. On the other hand, pairs can be excited without an
energy cost for densities $\rho_{\rm f}<1$. We further study the
single-particle (pair) correlation functions along the ladder,
obtaining exponential (power law) decay with distance, denoting gapped
(gapless) behavior for this excitation.  Finally, we
  found that power law decay of pair correlations are slower, often
  much slower, in the Creutz lattice than in the normal ladder. Note that
  Fig.~\ref{fig:Gpair}(b) shows that the decay exponent on the Creutz
  lattice is always less than $1/2$ while on the ladder,
  Fig.~\ref{fig:Gpair_ladder}(b) shows it always to be larger than
  $1/2$ and can even rise above $1$. With the Luttinger parameter,
  $K$, defined as $G(r)\sim 1/r^{(1/2K)}$, this gives $K>1$ for Creutz
  and $K<1$ for the ladder. It is known from
  bosonization\cite{Giamarchibook} that when $K<1/2$ ({\it i.e.} power
  decay exponent larger than $1$) the superfluid is unstable and can
  be localized even by a single impurity. In this sense, we say that
  superfluidity in the Creutz lattice is more robust.

\acknowledgments The authors acknowledge insightful discussions with
R. Scalettar. RM also acknowledges discussions with C. Cheng. RM is
supported by the National Natural Science Foundation of China (NSFC)
Grant No 11674021 and No. 11650110441; and NSAF-U1530401. GGB is partially
supported by the French government, through the UCAJEDI Investments in
the reference number ANR-15-IDEX-01. The computations were performed
in the Tianhe-2JK at the Beijing Computational Science Research Center
(CSRC) and with resources of the National Supercomputing Centre,
Singapore (\url{https://www.nscc.sg}). This research is supported by
the National Research Foundation, Prime Minister's Office, Singapore
and the Ministry of Education-Singapore under the Research Centres of
Excellence programme.

\textit{Note.} Upon completion of this manuscript, a preprint appeared
which tackled a similar problem~\cite{Tovmasyan2018}, also finding
formation of pair superfluidity away from half-filling. In addition to
obtaining similar results, here we also quantitatively connect the
results in the strongly interacting regime to a repulsive hardcore
boson model which is superfluid for any finite interactions. We also
show how those are connected to properties of the groundstate, as in
the single- and pair-correlation functions.

\bibliography{references}

\end{document}